\documentclass{kapproc} % Computer Modern font calls

\RequirePackage{graphicx}%
\RequirePackage{epsf}%
\input{psfig.sty}

\upperandlowercase
\let\footnote\savefootnote
\let\footnotetext\savefootnotetext 
\let\footnoterule\savefootnoterule 
\kluwerbib % will produce  bibliography 
\begin{document}

%------------ article title  ------------------->>

\articletitle{Gould's Belt to Starburst Galaxies:  The IMF of Extreme Star Formation}

%% optional, to supply a subtitle:
%\articlesubtitle{Spineto@50}

%% Supply a shorter version of the title for the running head:
\chaptitlerunninghead{Gould's Belt to Starburst Galaxies}

%------ author/affiliation choices -------------->>

%% Single author or several authors with same affiliation

 \author{M.R. Meyer, J. Greissl, M. Kenworthy, and D. McCarthy}
 \affil{Steward Observatory, The University of Arizona, Tucson, AZ 85721--0065 USA}
 \email{mmeyer@as.arizona.edu}

%% Multiple authors, multiple affiliations

%\author{First Author\altaffilmark{1}, Second Author\altaffilmark{2}, 
%         Third Author\altaffilmark{1,3}}

%\affil{\altaffilmark{1}Institute, Address, Country, \\ 
%\altaffilmark{2}Institute, Address, Country, \\
%\altaffilmark{3}Institute, Address, Country}

%\email{author1@add1,author2@add2,author3@add3}

% abstract
 \begin{abstract}
Recent results indicate the stellar initial mass function is not
a strong function of star--forming environment or ``initial conditions''
(e.g. Meyer et al. 2000).  Some studies suggest that a universal IMF may
extend to sub--stellar masses (see however Briceno et al. 2002).  Yet most
of this work is confined to star--forming environments within 1 kpc of the 
Sun.  In order to probe the universality of the IMF over a wider range
of parameter space (metalicity, ambient pressure, magnetic field strength) 
new techniques are required.  We begin by summarizing our approach to 
deriving the sub--stellar IMF down to the opacity--limit for fragmentation 
using NGC 1333 as an example.  Next, we
describe results from simulations using the observed point--spread function 
of the new 6.5m MMT adaptive optics system and examine the confusion--limited
sensitivity to low mass stars in rich star--forming clusters out to 0.5 Mpc.
We also present preliminary results from observations with this system of
the W51 star--forming complex.  Finally, we outline a new technique
to estimate the ratio of high to low mass stars in unresolved stellar
populations, such as the massive star clusters observed in interacting
galaxies (e.g. Mengel et al. 2002).  While evidence for variations in 
the IMF remains inconclusive, new studies are required to rule them 
out and determine whether or not the IMF is universal over the range
of parameter space relevant to star--forming galaxies over cosmic time. 
 \end{abstract}

%------------ body of article ------------------->>
\section{Introduction}

Any predictive theory of star formation must explain the observed 
shape of the field star IMF (e.g. Chabrier, 2003) and any deviations
from it (e.g. Figer et al. 1999) as a function of physical properties. 
There are a few critical scales in star formation that
one might expect {\it should} depend on local conditions such 
as super--critical mass to overwhelm magnetic support 
(e.g. Shu et al. this volume), or the Jean's mass (e.g. Larson, this volume). 
Curiously, the stellar IMF observed
toward a number of star--forming regions within 1 kpc of the Sun
diplay no evidence for a strong dependence on ``initial conditions''
and are consistent with having been drawn from the same IMF
that characterizes the field (Meyer et al. 2000).  For this reason, 
we are forced to expand the range of parameter space
in which we search for variations in the IMF.  There are two 
obvious avenues for further exploration:  1)  probing the sub--stellar
IMF in nearby regions down to and below the expected minimum mass 
for opacity--limited fragmentation (e.g. 0.01 M$_{\odot}$ Spitzer, 1978); 
and 2) more {\it extreme} star formation over a
broader span of physical conditions such as metalicity, stellar 
density, galactic environment. 

\section{Probing to the End of the IMF}

Advances in optical and 
infrared instrumentation on large ground--based telescope, 
as well as the Hubble Space Telescope have
led to a number of IMF determinations that probe well into the 
brown dwarf regime. Several groups have derived sub--stellar IMF slopes
in clusters that are flat, or falling, in log--mass units 
$dN/dlogm = m^{-\Gamma}$ with $\Gamma = 0$ to 1 
(e.g. Bouvier et al. 1998; Hillenbrand and Carpenter, 2000; Najita et al. 
2000).  While most regions appear to be consistent 
with each other, the Taurus dark cloud 
appears to have a statistically significant 
dirth of brown dwarfs compared to the Trapezium 
cluster (Briceno et al. 2002).  Whether this is due to Taurus being
deficient in brown dwarfs or the Trapezium being overabundant, is 
still not clear (Luhman, this volume). 

A standard approach applied to the study of many young embedded clusters
involves deep infrared imaging surveys along with follow-up infrared 
spectroscopy for as many cluster members as is practical (e.g. Greene 
and Meyer 1995; Luhman \& Rieke 1998).  
More recently, it has been 
possible to obtain IR spectra for low luminosity sources enabling astronomers 
to identify young brown dwarf candidates in large numbers 
(e.g. Luhman 1999; Wilking et al. 1999).   One concern of these
early studies was the use of photospheric absorption features to derive
spectral types which might also be effected by surface gravity 
(lower in pre--main sequence ``sub--giants'' compared to high 
gravity dwarf standards).  In a recent study of brown dwarf candidates
in NGC 1333, Wilking et al. (2004) test whether their spectrophotometric
reddening--independent $H_2O$ K--band index is effected by gravity.  
They find that any potential bias introduced by comparing PMS
brown dwarf candidate spectra to dwarf star standards is of order
the error inherent in the classification scheme.   This is consistent
with the work of Gorlova et al. (2003) who also find that
$H_2O$ in the J--band depends weakly on surface gravity, making it
a good temperature indicator for young late--type stars.  However, 
Gorlova et al. did find that the 1.25 $\mu$m KI feature, 
along with the 2.2 $\mu$m NaI feature, is sensitive to surface gravity 
providing a possible tool to distinguish foreground and background
stars from ``sub--giant'' PMS cluster members.  We are
currently following up this work with an approved Cycle \# 13 program 
on HST using NICMOS in slitless grism mode
to obtain 1.1--1.9 $\mu$m spectra of brown dwarf candidates in NGC 1333.  
Our goal is to probe the IMF down below 10 M$_{JUP}$ (0.01 M$_{\odot}$)
in this intermediate density cluster (between Taurus and the Trapezium) 
to see if there is a correlation with the slope of the sub--stellar IMF. 

Another curiousity observed in mass distributions that result from 
star formation processes, is the brown dwarf desert in the companion mass
ratio distribution (CMRD).  It is fairly well established from radial velocity
surveys as well as seeing--limited common proper motion surveys 
(Udry et al. 2001; Hinz et al. 2001) that there is a dirth of brown dwarf 
companions compared to extrapolation of the 
CMRD of stellar or planetary companions. 
In fact, the CMRD from 0.1--1.0 M$_{\odot}$ surrounding solar--type stars 
(Duquennoy \& Mayor, 1991; DM91) is 
consistent with having been drawn from a field star IMF, 
in stark contrast to the frequency of brown dwarf companions to sun--like
stars compared to the sub--stellar IMF in the field (Ried et al. 1999).
What about the companion mass ratio distribution surrounding young PMS stars?
It has been known for some time that the binary frequency of T Tauri 
stars is high, particularly in low density star--forming regions such
as Taurus--Auriga (Ghez et al. 1993; Leinert et al. 1993).    However, 
it is only recently that researchers have begun to obtain multi--color
infrared photometry and spatially--resolved IR spectra needed to derive
mass estimates for faint companions to T Tauri stars.  In fact, detailed
comparisons of the CMRD between young clusters and the field are lacking.

In a NICMOS HST study of NGC 2024, Liu et al. (2003) describe how one 
must carefully consider completeness limits for faint companions
as a function of angular separation and then compare the observations
to results for field stars only over the appropriate mass ratios as 
a function of radius from the primary.  They also consider what one
might expect as a function of cluster age if soft--binaries are 
preferentially disrupted faster in higher density regions.  Taking 
these effects into account, and combining their data with published
results, they find a significant inverse correlation between binary frequency 
and stellar density.  Fundamental 
survey work still needs to be done on the field star population as well 
as young star--forming regions, particular with the recent advances in 
adaptive optics on large telescopes.  It will be important to know in this
inherently three--dimensional observational problem whether: 
1) the CMRD is a function of separation; and 2) whether
the CMRD or the integrated period distribution is a function of 
primary mass.  Although strong observational evidence to support 
the ejection hypothesis (e.g. Reipurth and Clarke, 2001) 
for the formation of free--floating brown dwarfs is lacking
(e.g. Muench et al. 2002)
it is curious to note that the one region deficient
in brown dwarfs in the system IMF (Taurus) is also the lowest
density star--forming region with the highest observed binary fraction. 

What does the future hold in studying the sub--stellar IMF in 
star--forming regions?  Efforts continue to push down below 
the minimum mass for opacity--limited fragmentation in nearby
young clusters.  If there is a lack of objects below some cutoff
mass, it will be hard to miss observationally, and particularly
exciting if it depends on local conditions.   Peale (1999)
points out that if there are significant numbers of brown dwarfs below
0.01 M$_{\odot}$ in young clusters, it may suggest an IMF that
varies over cosmic time as recent MACHO results rule out
significant mass in the halo in objects $<$ 0.01 M$_{\odot}$.  One must be 
cautious in identifying extremely low mass objects in young clusters
from photometry alone as there can be significant contamination
in color--magnitude diagrams in deep surveys.  We believe
that spectroscopic follow--up is essential and spectroscopic
diagnostics of surface gravity may be required in order to remove
old cool field dwarfs from young cluster samples (Burgasser et al. 
2004).  We are working to identify a set of surface gravity
diagnostics which could be used in filter--photometric imaging
surveys to rapidly identify large samples of young late--type objects
in star--forming regions.  This effort complements recent work developing 
filter--photometric temperature diagnostics for very cool photospheres
(e.g. Mainzer et al. 2003). 

\section{Star Formation Extreme}

How does star formation in the solar neighborhood compare
to global star formation in the Milky Way or even other
galaxies?  If one adds up the total number of high mass
stars along the Gould's Belt within 1 kpc of the Sun 
(de Zeeuw et al. 1999) and assumes a field star IMF, 
then the integrated total mass of the system is about
3 $\times$ 10$^5$ M$_{\odot}$, comparable to a single 
``event'' such as the R 136 cluster region in the LMC.  
Is the IMF in extreme star--forming regions such as R 136
(low metalicity, disturbed galactic morphology, high stellar
density), or the galactic center (high metalicity, 
ambient pressure, and magnetic flux density) 
and comparable to that observed
in the solar neighborhood?  Can we probe further claims
of unusual IMFs indirectly inferred from observations of 
starburst galaxies?   While we may not be able to derive
detailed distributions of the numbers of individual objects
as a function of mass in these distant 
stellar systems, we can apply similar techniques to 
those described above to place limits on the ratio of high
to low mass stars in these unusual regions that represent
accessible extreme star--forming environments. 

\subsection{The Limits of Confusion}

In principle, one would like to conduct observational 
``experiments'' where one could compare differentially
results from observations of star--forming regions varying
one parameter at a time in a controlled way.  However, such 
studies are thwarted by the limits one can achieve in sensitivity
and spatial resolution for rare targets which tend to lie
at the greatest distances.  In particular, studies of 
the IMF in rich stellar fields are often 
{\it confusion--limited} rather than sensitivity--limited.
For example, Andersen et al. (in preparation) find that NICMOS
HST observations of the IMF toward R 136 in the LMC are 
confusion--limited at $>$ 0.5 M$_{\odot}$.  In this case, 
the spatial resolution afforded by large--ground based telescopes
can help, but only if the point--spread function (PSF) 
is smooth and stable (Stolte et al.  2003).  

In order to see how well one can do with currently operating
adaptive optics (AO) systems on 6--10 meter telescopes, we performed a series of
simulations using the observed PSF of the 6.5m MMT with the
new adaptive secondary mirror (Close et al. 2003).  We considered
the Trapezium cluster observations of Hillenbrand and Carpenter
(2000) as input, and simulated what the Trapezium would look
like at 5 kpc, 25 kpc, 50 kpc, and 0.5 Mpc.  As the cluster
gets farther away, the residual uncorrected halo of the bright
Trapezium stars dominates the background at progressively larger
radii.  At 5 kpc, one still reaches the sky background limit at 0.2 
parsecs enabling one to easily detect objects at the sub--stellar
boundary in reasonable integration times of a few hours.  At 25 kpc
(Figure 2), 
one reaches the sky background limit $\times$ 5 farther out still 
enabling one to measure a significant number of stars at 2.5 R$_{core}$
over the entire stellar mass range.  At the core radius of the 
cluster inside of which most of the stellar population
is located, one loses $\Delta H$ = 6$^m$ of sensitivity due to 
the background of uncorrected halo light from the bright stars. 
At 50 kpc, one does not hit the sky background limit until 4 R$_{core}$
and the sensitivity at the core radius is almost nine magnitudes
worse than the natural background.  At 0.5 Mpc, the 
cluster is competely unresolved.  We conclude that 
while it will be relatively straightforward to characterize the 
IMF down to the hydrogen burning limit for any Trapezium in the 
Milky Way galaxy with existing AO systems on 6--10 meter telescopes, 
achieving similar results in nearby galaxies will require higher
spatial resolution afforded by the next generation of large telescopes
(e.g. the Large Binocular Telescope). 

\begin{figure}[ht]
\centerline{\psfig{file=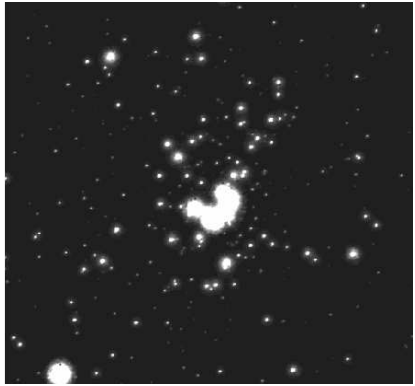,height=2in}}
\caption{Simulations of the Trapezium cluster projected to a distance 
of 25 kpc as it would appear observed with the MMT adaptive optics system. 
These observations would reach the natural background limit outside
of 2.5 core radii enabling determination of the IMF over the full 
range of stellar masses. 
} 
\end{figure}

Figer et al. (1999; this volume) 
have presented intriguing observations that suggest
the IMF in the inner galaxy Arches cluster is flatter than the
field star IMF over the mass range observed (2--20 M$_{\odot}$).  
In order to test whether or not the ratio of high to low mass stars in 
the Milky Way depends on metalicity, ambient ISM pressure, 
and/or magnetic field strength, 
we have begun a program using the ARIES camera in tandem with 
the MMT--AO system to image massive star forming clusters at 
high spatial resolution and sensitivity. 
Our first target is the distant luminous UCHII region W51 which contains 
over 130 stars $>$ 
10 M$_{\odot}$ (Okumura et al. 2000).  Preliminary analysis of observations 
obtained in May, 2004 (Figure 2), indicate that we can reach the hydrogen 
burning limit in this cluster through A$_v$ = 30$^m$ of extinction
(K $<$ 21$^m$) 
in one hour of on--source integration time.  We are currently exploring whether
narrow--band IR imaging at R=100 in 6--8 filters can provide the 
necessary spectrophotometric information to estimate rough temperatures
and surface gravities needed to distinguish foreground and background
stars along this complex line of sight (Meyer et al. 1998). 

\begin{figure}[ht]
\centerline{\psfig{file=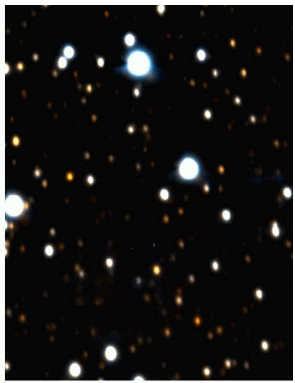,height=2in}}
\caption{Observations of W51 in the H- and K-bands using the ARIES IR Camera
and the MMT-AO system.  Coadding 20 minutes of data resulted in 
images with 0.1'' FWHM in the K--band. Preliminary results suggest 
that we will be able to reach the hydrogen--burning limit in the
cluster in one hour of on--source integration time in the K--band.}
\end{figure}

\subsection{Unresolved Super--Star Clusters}

Some of the longest lasting claims of unusual IMFs come
from observations of starburst galaxies with star--formation
rates orders of magnitude higher than normal galaxies.
Super--star clusters, recognized as
globular cluster analogues in the early stages of formation
(O'Connell et al. 1994) 
have been found frequently within interacting
starbursting galaxies.  Mengel et al. (2002) 
used dynamical mass estimates from velocity dispersions to find 
significant variations in inferred mass--to--light ratios
which they interpret as evidence for IMF variations from
cluster to cluster in NGC 4038/39.  Is there any way we could
directly detect the low mass stars in these super--star clusters
and constrain the ratio of high to low mass stars? 

We have explored the contribution to the integrated K--band light
from very young low mass stars in super--star clusters $<$ 10 Myr old 
(Meyer and Greissl, 2004). 
While a main sequence population is dominated by the light from the highest mass
stars because of the steepness of the main sequence mass--luminosity
relation, in the PMS this M--L relation is much flatter perhaps
rendering the low mass stars visible for a short time in a forming
super--star cluster.  
Combining an assumed IMF (Salpeter, 1955 or Chabrier, 2003) with 
PMS evolutionary models (Siess et al. 2000) and predictions of 
Starburst99 for the main sequence, post--main sequence, and nebular
contributions to the integrated light, we find that between 7--12 \% 
of the K--band flux comes from young late--type stars depending
on which IMF is assumed.  Coupling these luminosity estimates
with spectra appropriate for each object in the synthetic 
luminosity function, we have created model integrated spectra
for the combined stellar population (Figure 3).  Because these
late--type stars are overluminous in the PMS, and because their
spectra are dramatically different, with strong broad 
absorption bands from CO and other molecules, we predict that 
in a super--star cluster $<$ 10 Myr (before the formation of the
first red super--giants dominate the K--band light of the cluster)
one might detect directly the integrated light of the low mass
stars as a 2--4 \% absorption feature against the continuum.
If one could use observations at centimeter and other wavelengths to 
models the contribution of nebular free--free emission to the 
2 $\mu$m continuum, the strength of the CO feature from late--type
stars could provide a direct constraint on the ratio of high to 
low mass stars in these super--star clusters.  The high signal to 
noise spectra ($>$ 100) needed can be obtained at R=2000 with NIRSpec
on the Keck telescope in approximately five hours of integration
time for the brightest candidate SSCs in NGC 4038/39 that are 
thought to be $<$ 10 Myr (Frogel et al. 2003).  

It must be cautioned that we 
have ignored some aspects of the problem in these simulations.  
Near--IR excess emission from disks surrounding the young
low mass stars might {\it dilute} the expected absorption strength of
the photospheric features in the IR (Meyer et al. 1997). 
However, because PMS stars have sub--giant surface gravities, 
we are also {\it underestimating} the strength of CO feature 
which is surface gravity sensitive (Kleinmann and Hall, 1986). 
We do not argue that these simulations are a perfect representation 
of what one should expect from the integrated light of super--star clusters.  
Nevertheless, we draw inspiration from Salpeter (2002), who, while musing about 
the longevity of his 1955 paper, suggested that it 
is not important for all the assumptions in a calculation be correct, 
but that the largest uncertainties should tend to cancel! 

\begin{figure}[ht]
\centerline{\psfig{file=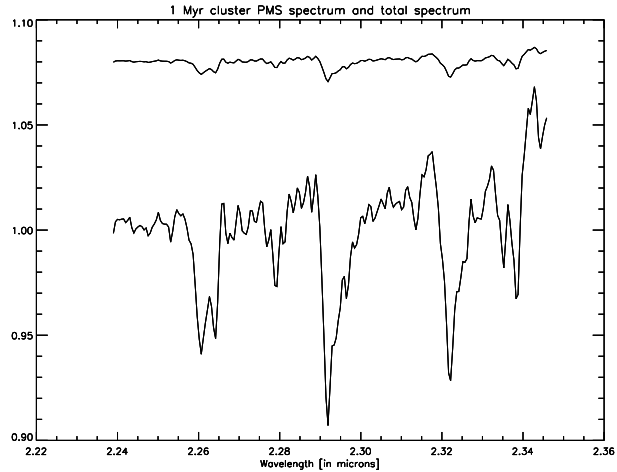,height=2.5in,angle=90}}
\caption{The total integrated K--band 
spectrum of a 1 Myr old 10$^6$ M$_{\odot}$ unresolved
super--star cluster (top) and the associated spectrum of 
the PMS stars only (bottom).  In a few hours of integration
time on a 6--10 meter telescope, we should be able to detect
the signature of the late--type low mass stars (2--4 \% absorption
features against the continuum) in candidate 
super--star clusters in NGC 4038/39.}
\end{figure}

\section{Conclusions}  

Our conclusions can be briefly summarized at follows:  1) the stellar
IMF observed in nearby star--forming regions is broadly consistent with
the field star IMF and does not vary strongly with local conditions 
(Meyer et al. 2000); 2) sub--stellar objects contribute less than 10 \% 
of the dynamical mass of young stellar populations (Wilking et al. 
2004) and this fraction might vary from region to region (Briceno et al. 2002); 
3) there may be a connection between the binary fraction in a young
cluster, its stellar density, and the observed system IMF (Liu 
et al. 2003); 4) spectroscopic estimates of temperature and surface 
gravity are needed in order to survey down below the expected minimum 
mass for fragmentation in nearby molecular clouds (Gorlova et al. 
2004); 5) while existing AO systems on 6--10m telescopes should
enable us to probe down to the hydrogen burning limit throughout
the Milky Way galaxy (e.g. W51), it will be extremely difficult to reach this
limit even in the local group without higher spatial resolution
afforded by the next generation of very large telescopes such as 
the LBT; and 6) it may be possible to detect directly the low mass
stellar component in the high SNR spectrum of an integrated super--cluster 
in nearby interacting galaxies placing constraints on the ratio of 
high to low mass stars (Meyer and Greissl, 2004).

\begin{acknowledgments}
We would like to thank the meeting organizers for providing such a 
stimulating get--together in such a lovely place 
where we could share ideas on this work with such generous and
talented colleagues. 
We congratulate Prof. Salpeter on his lifetime of important achievements
in astrophysics and wish him many more years of good health and happiness. 
This work was generously supported by a Cottrell Scholar's Award to MRM 
from the Research Corporation and NASA grant HST13-9846. 
\end{acknowledgments}

\end{document}